\documentclass[acmsmall,nonacm]{acmart}

\usepackage{graphicx} 
\usepackage{color,soul}

\title{Expanding Specification Capabilities of a Gradual Verifier with Pure Functions}
\author{Doruk Alp Mutlu}
\affiliation{
    \institution{Michigan State University}
    \country{USA}
}
\email{mutludor@msu.edu}

\usepackage{booktabs}   
\usepackage{caption} 
\usepackage{wrapfig} 
\usepackage{subcaption} 
\usepackage{setspace}
\usepackage{newunicodechar}
\usepackage{listings}
\usepackage{placeins}
\usepackage{xspace}
\usepackage{plstx, listproc}
\usepackage{svg}
\newunicodechar{≠}{=\llap{/}}
\newunicodechar{≔}{:\raisebox{-0.18ex}[\height][\depth]{=}}
\newunicodechar{≤}{\rlap{\begingroup \fontsize{13pt}{13pt}\selectfont \raisebox{-0.58ex}[\height][\depth]{-} \endgroup}\shifttext{0.15ex}{\raisebox{0.18ex}[\height][\depth]{<}}}
\newunicodechar{≥}{\rlap{\begingroup \fontsize{13pt}{13pt}\selectfont \raisebox{-0.58ex}[\height][\depth]{-} \endgroup}\shifttext{0.15ex}{\raisebox{0.18ex}[\height][\depth]{>}}}
\newunicodechar{∧}{\shifttext{1ex}{\begingroup \fontsize{8pt}{8pt}\selectfont {\shifttext{0.4ex}{/}\shifttext{-0.4ex}{\textbackslash}} \endgroup}}
\newunicodechar{∨}{\shifttext{1ex}{\begingroup \fontsize{8pt}{8pt}\selectfont {\shifttext{0.4ex}{\textbackslash}\shifttext{-0.4ex}{/}} \endgroup}}

\makeatletter
\newcommand*{\shifttext}[2]{%
	\settowidth{\@tempdima}{#2}%
	\makebox[\@tempdima]{\hspace*{#1}#2}%
}
\makeatother

\usepackage[leftmargin=3em,vskip=0em]{quoting}
  
\definecolor{light-gray}{gray}{0.87}

\definecolor{gray}{gray}{0.75}

\definecolor{drk-gray}{RGB}{169,169,169}

\definecolor{light-purple}{RGB}{229,204,255}

\definecolor{light-yellow}{RGB}{255,228,181}
\newcommand{\lightyellow}[1]{\colorbox{light-yellow}{#1}}
\newcommand{\mlightyellow}[1]{\mcolor{light-yellow}{#1}}

\definecolor{light-blue}{RGB}{189,215,238}

\newcommand{\mlightblue}[1]{\mcolor{light-blue}{#1}}

\definecolor{light-green}{RGB}{152,251,152}

\definecolor{light-red}{RGB}{255,204,204}

\definecolor{rred}{RGB}{255,153,153}

\definecolor{light-pink}{RGB}{224,191,184}

\definecolor{light-orange}{RGB}{255,204,153}

\definecolor{neon-blue}{RGB}{153,255,255}

\definecolor{neon-yellow}{RGB}{255,255,153}

\newcommand{\mcolor}[2]{\colorbox{#1}{$\displaystyle #2$}}

\newcommand{\grad}[1]{\widetilde{#1}}

\newcommand{\gradTT}[1]{\ensuremath{\grad{#1}}}

\newcommand{\gphi}{\grad{\phi}}

\newcommand{\std}{\textrm}
\newcommand{\ttt}[1]{\textup{\texttt{\small{#1}}}}
\newcommand{\disableTttResize}[0]{\renewcommand{\small}[1]{##1}}

\newcommand{\predicate}[1]{\textup{\textmd{\textsf{#1}}}}

\newcommand{\gviper}{\std{Gradual Viper}\xspace}

\newcommand{\qm}{\ttt{?}}

\newcommand{\phiCons}[2]{{#1}\:\&\&\:{#2}}

\newcommand{\withqm}[1]{\phiCons{\qm}{\ensuremath{#1}}}

\newcommand{\sCall}[3]{#1~\ttt{≔}~#2\ttt{(}#3\ttt{)}}

\newcommand{\produce}[4]{\predicate{produce}\ttt{(}#1,~#2,~#3,~#4\ttt{)}}

\newcommand{\hconsume}[3]{\lightyellow{\predicate{consume}}\ttt{(}#1,~#2,~#3\ttt{)}}

\newcommand{\eval}[3]{\predicate{eval}\ttt{(}#1,~#2,~#3\ttt{)}}
\newcommand{\heval}[3]{\lightyellow{\predicate{eval}}\ttt{(}#1,~#2,~#3\ttt{)}}

\newcommand{\verify}[1]{\predicate{verify}\ttt{(}#1\ttt{)}}
\newcommand{\wellformed}[4]{\predicate{well-formed}\ttt{(}#1,~#2,~#3,~#4\ttt{)}}

\newcommand{\wellformede}[3]{\predicate{well-formed-e}\ttt{(}#1,~#2,~#3\ttt{)}}

\newcommand{\hwellformed}[4]{\lightyellow{\predicate{well-formed}}\ttt{(}#1,~#2,~#3,~#4\ttt{)}}

\newcommand{\hwellformede}[3]{\lightyellow{\predicate{well-formed-e}}\ttt{(}#1,~#2,~#3\ttt{)}}

\newcommand{\rchecks}{\mathcal{R}}

\newcommand{\origin}{\ttt{origin}}
\newcommand{\none}{\ttt{none}}

\newcommand{\success}{\predicate{success}\ttt{(}\ttt{)}}

\newcommand{\fresh}{\predicate{fresh}}
\newcommand{\assign}{~\ttt{≔}~}
\newcommand{\oh}{h_{\qm}}
\newcommand{\isimp}{\ttt{isImprecise}}

\usepackage{pdfpages}
\usepackage{multirow}
\usepackage{multicol}
\usepackage{array} 
\usepackage{savesym} 

\usepackage{tikz}

\definecolor{legend_green}{HTML}{33A02C}
\definecolor{legend_purple}{HTML}{6A3D9A}
\definecolor{legend_red}{HTML}{DE2D26}
\definecolor{legend_yellow}{HTML}{FFD700}
\definecolor{legend_green_light}{HTML}{c8e4c4}
\definecolor{legend_purple_light}{HTML}{c4c4e8}

\usepackage{minted}
\usepackage{algorithmicx}
\usepackage{algpseudocode,algorithm,algorithmicx}

\algrenewcommand\algorithmicrequire{\textbf{Precondition:}}
\algrenewcommand\algorithmicensure{\textbf{Postcondition:}}

\newlength\myindent
\begin{document}
\maketitle
\vspace*{-0.30cm}
Gradual verification soundly combines static checking and dynamic checking to provide an incremental approach for software verification. With gradual verification, programs can be partially specified first, and then the full specification of a program can be achieved in incremental steps. The first and only practicable gradual verifier based on symbolic execution, Gradual C0, supports recursive heap data structures.  Despite recent efforts to improve the expressivity of Gradual C0's specification language, Gradual C0's specification language is still limited in its capabilities for complex expressions. This work explores an extension to Gradual C0's design with a common construct supported by many static verification tools, pure functions, which both extend the specification capabilities of Gradual C0 and increase the ease of encoding observer methods in Gradual C0. Our approach addresses the technical challenges related to the axiomatisation of pure functions with imprecise specifications.

\section{Introduction}
Static verification requires complete specifications to ensure that a program adheres to its specifications. Even though static verification can guarantee the correctness of a program given the specifications, this process burdens users by requiring detailed specifications. On the other hand, dynamic verification supports partial specifications of program properties through run-time checks. However, this comes at a significant cost of run-time overheads, and dynamic verification is limited to verifying a program during run-time. Combining these two techniques, gradual verification,  first introduced by Bader et al.\cite{bader2018gradual, wise2020gradual}, offers a smooth transition between static verification and dynamic verification. Gradual verification relies on imprecise specifications (conjoining precise formulas with ?) to make this transition possible. In the presence of imprecise specifications, run-time checks are inserted in the original program where static information is not sufficient for static verification. 

The current practicable implementation of a gradual verifier, Gradual C0\cite{divincenzo2024toplas}, is based on symbolic execution and built on top of the Viper verification infrastructure\cite{MuellerSchwerhoffSummers16}. Differing from various front-end modules such as VerCors\cite{vercors2016}, which rely on the Viper verification infrastructure, Viper's symbolic execution engine and intermediate language are modified to handle the imprecision introduced by imprecise specifications in gradual verification. Both Viper and by consequence Gradual C0 support the implicit dynamic frames logic for reasoning about heap manipulating programs, and recursive abstract predicates to reason about recursive heap data structures. 

One of the current limitations of Gradual C0 is the expressivity of its specification language. Recent work by Gupta \cite{gupta2025} introduced and implemented the design of unfolding expressions in the gradual setting for Gradual C0, which supports more intuitive specifications for recursive heap data structures. Pure functions, which are side-effect free constructs, enable intuitive specifications for recursive heap data structures as well. Pure functions enable this by making it possible to encode side-effect free observer methods and expressing refinements of existing predicates\cite{MuellerSchwerhoffSummers16}. Also, pure functions often use unfolding expressions as well. A common example is unfolding a predicate in the function body to make the permissions required in the recursive function call available. One important detail here is that both the unfolding expressions and pure functions are side-effect free, meaning both can be used in specifications in various forms. By extending the design of gradual verification with pure functions, users can write specifications for recursive heap data structures more easily.

\begin{wrapfigure}{l}{0.5\textwidth} 
    \vspace{0pt}
    \hspace{0pt} 
    \begin{minipage}[t]{\linewidth} 
        \centering
        \begin{lstlisting}
int pureFactorial(int n)
  //@ pure;
  //@ requires n >= 0;
{ 
  return n == 0 ? 1 : n * pureFactorial(n-1);
}

int iterativeFactorial(int n)
  //@ requires n >= 0;
  //@ ensures \result == pureFactorial(n);
{
  int result = 1;
  int i = 1;

  while (i <= n)
    //@ loop_invariant 1 <= i && i <= n + 1;
    /*@ loop_invariant result ==
                pureFactorial(i - 1); @*/
  {
    result = result * i;
    i = i + 1;
  }

  return result;
}
        \end{lstlisting}
        \caption{Verifying an iterative factorial method in \\Gradual C0 using a pure function}
    \end{minipage}
\end{wrapfigure}

An important motivation to support pure functions in Gradual C0 is to enable the specification of methods which are not possible to specify with the current constructs in Gradual C0. This is common with iterative methods, such as computing Fibonacci numbers or summations over a range. As given in Figure 1, we focus on the example of factorial computation. First, a recursive pure function (denoted with the pure tag) is given, which computes the factorial of a non-negative integer $n$. Then, an iterative method with the same goal is given. In this case, the given pure function enables the verification of the method. One critical detail here is that methods, which are impure, cannot be used in specifications, but pure functions can be used in specifications through function calls.



\section{Approach}
The symbolic execution engine for Gradual C0, called Gradual Viper, relies on the four major functions of Viper modified for gradual verification. The four functions are eval, produce, consume, and exec. Eval is used for evaluating expressions and exec is used for executing statements. Producing a formula adds permissions, and constraints, while consuming a formula asserts constraints and removes permissions. As Gradual C0 extends the Viper verification infrastructure, pure functions in Gradual C0 are designed similarly to the pure functions in Viper. To better understand how our approach builds on the theory of verification of functions in Viper, we first explore how functions are verified in Viper.

In Viper, before verification of any construct, functions have their body and postcondition checked with respect to the permissions and constraints provided in the precondition. This stage is also known as the stage for checking the well-definedness of functions. In this stage, axioms related to the functions are generated and emitted to the underlying SMT solver; this process is called axiomatisation. When the function calls need to be evaluated later, the axioms from the SMT solver are used. This prevents re-evaluation of functions in each function call. Emitted axioms reason about the values of heap locations through function snapshots, which are binary tree representations of the current values on the heap needed by the functions. During the evaluation of the function calls, function snapshots are computed during the consumption of function preconditions and passed to the symbolic function application\cite{schwerhoff2016advancing}. 


The design of pure functions for Gradual C0 modifies both the axiomatisation algorithm and the eval rule for functions in Viper. The main challenge with the verification of pure functions in Gradual C0 arises from allowing pure function specifications to be imprecise. In the well-definedness stage of function verification, the function precondition is produced, during which the initial function snapshot is created. The mappings between the snapshot components and heap locations are recorded. During the evaluation of the function body and postcondition, heap accesses are related to the components of the function snapshot using the previously found mappings. However, the function precondition can be imprecise, which causes the function snapshot to be incomplete. An incomplete function snapshot would result in a failure of the axiomatisation process. As a solution, when a precondition is imprecise and a heap access from the precondition, body, or postcondition of a function cannot be mapped to a component of the function snapshot, we attempt to extend the function precondition and the function snapshot with the missing permissions. This is done with respect to the branch conditions. If the function snapshot extension attempt results in failure, the user is informed of the reason. The major known reason is discussed later. For the later evaluations of the function in the function calls, we require the permissions related to the extended component of the function snapshot to be consumed in addition to the initial function precondition.



\begin{wrapfigure}{r}{0.5\textwidth} 
    \vspace{-17pt}
    \hspace{8pt} 
    \begin{minipage}[t]{\linewidth} 
        \centering
        \begin{lstlisting}
struct Node { int val; };
typedef struct Node Node;
int double(bool b, Node* x, Node* y, Node* z)
  //@ pure;
  //@ requires ? && acc(y->val);
  //@ requires b ? z == x : z == y;
{
  return b ? x->val + z->val : y->val + z->val;
}
        \end{lstlisting}
        \caption{Pure function example in Gradual C0 with\\ imprecise specifications}
    \end{minipage}
\end{wrapfigure}

To illustrate how the previously defined axiomatisation strategy for Gradual C0 works, we now consider the example in Figure 2, which is a modified version of an example from the foundational work on function axiomatisation in Viper\cite{schwerhoff2016advancing}. In the original example, \lstinline|acc(x->val)| is provided instead of ?. During well-definedness check for this pure function in Gradual C0, when the body of the pure function is evaluated, the heap access x->val fails to find the corresponding snapshot. However, as the precondition is imprecise, we can optimistically assume the permission for x->val is provided in the precondition. Then, we can transform (assume) the third line of the pure function's precondition is \lstinline|b ? acc(x->val) && z == x : z == y| which is semantically equivalent to the full specification. The function snapshot is extended, and the accessibility predicate \lstinline|acc(x->val)| needs to be consumed during the eval of any function application.

The approach described above has shortcomings with imprecise recursive predicates. Attempting to make optimistic assumptions about the accessibility predicates used in imprecise recursive predicates would require a permanent change in the shape of the predicate snapshots. This case can be understood as follows: the client method folds the predicate P, but the function snapshot required by the pure function may have a different shape for predicate P rather than the shape of predicate P folded by the client method. This situation occurs because of the optimistic assumptions. Therefore, imprecise recursive predicates cannot be used with the pure functions in Gradual C0. Functions with such preconditions are rejected by equi-recursively checking the preconditions. Postconditions do not need to be checked, as the postconditions must be framed by the preconditions. 

Another important part of our design for pure functions in Gradual C0 is evaluating the function calls. The rule used for eval (Figure 5, in Appendix) is modified to collect the run-time checks generated by consuming the function precondition. Also, $is-imp$ used in the rule checks if the pure function postcondition is imprecise. If the function postcondition is imprecise, we make the state imprecise. This exception is needed as the function postconditions are used in generated axioms, and imprecision is handled in symbolic execution at a level higher than the interactions with the SMT solver. 

\section{Conclusion}

Extending Gradual C0 with the pure functions allows the development of more straightforward specifications. This extension involves introducing a new strategy for axiomatising pure functions in a gradual setting as well as modifying the existing formal rule for functions in Viper for Gradual C0. Future work will explore implementing the gradual parts of this approach in Gradual C0 and soundness proof for pure functions based on the formalization introduced by Zimmerman et al. \cite{zimmerman2024sound}.

\newpage

\bibliography{references}
\clearpage

\section*{A \;Appendix}

\label{sec:figure-draft}

\begin{figure}[htbp]
\footnotesize\ttfamily
\disableTttResize
\hspace{-29.6em} $ \verify{decl} : MDECL \cup PDECL \longrightarrow RESULT$
\begin{alignat*}{2}  
  &\verify{\mathtt{function}~ \ fun(\overline{x:T}):~ T_r} = \\
    &&&\hspace{-5em}\hwellformed{\sigma_0\{ \gamma \assign \sigma_0.\gamma[\overline{x \mapsto \fresh }]\}}{func_{pre}}{\fresh}{(\lambda~ \sigma_1 ~.~ \\
    &&&\hspace{-5em}\qquad \hwellformede{\sigma_1\{\gamma \assign \sigma_1 . \gamma [\overline{\textup{\textmd{\textsf{result}}} \mapsto \fresh }] \}}{func_{post}}{ \\
    &&&\hspace{-5em}\qquad\quad (\lambda~ \_ ~.~ \success )} \\
    &&&\hspace{-5em}\qquad \wedge~ \\
    &&&\hspace{-5em}\qquad \heval{\sigma_1}{func_{body}}{(\lambda~ \sigma_2, body .  \\
    &&&\hspace{-5em}\qquad\quad \hconsume{\sigma_2\{\gamma \assign \sigma_2.\gamma[\overline{\textup{\textmd{\textsf{result}}} \mapsto \fresh }] \}}{func_{post}}{(\lambda~ \sigma_3,~ \_ ~.~ \\
    &&&\hspace{-5em}\qquad\qquad ~;~ \success )})})}\\
\end{alignat*}

\fbox{\begin{tabular}{llll}
\textcolor{light-yellow}{$\blacksquare$} & \small{Handles imprecision}
\end{tabular}}

\caption{Rule defining a valid pure function in a \gviper program}
\label{fig:figure-draft}
\end{figure}
\begin{figure}[!ht]
\scriptsize\ttfamily
\disableTttResize
\hspace{2.8em} 
\begin{alignat*}{2}
    \wellformed{\sigma}{\gphi}{\delta}{Q} & : \Sigma \longrightarrow \gradTT{FORMULA} \longrightarrow SNAPSHOT \longrightarrow (\Sigma \longrightarrow RESULT) \longrightarrow RESULT \\
    \hspace{-10em}\wellformede{\sigma}{\gphi}{Q} & : \Sigma \longrightarrow \gradTT{FORMULA} \longrightarrow (\Sigma \longrightarrow RESULT) \longrightarrow RESULT
\end{alignat*}
\hspace{-5em}
\begin{alignat*}{2}
    & \wellformed{\sigma_1}{\gphi}{\delta}{Q}
    &&= \produce{\sigma_1}{\gphi}{\delta}{(\lambda~ \sigma_2 ~.~\\
    &&&\qquad \produce{\sigma_1\{~ \pi \assign \sigma_2.\pi ~\}}{\gphi}{\delta}{Q})} \\
    & \wellformede{\sigma_1}{\phi}{Q}
    &&= \eval{\sigma_1\{\isimp \assign \text{false}\}}{\phi}{Q} \\
    & \wellformede{\sigma_1}{\withqm{\phi}}{Q}
    &&= \eval{\sigma_1\{\isimp \assign \text{true}\}}{\phi}{Q}
\end{alignat*}
\caption{Well-formed formula function definition}
\label{fig:wellformed-func}
\end{figure}

In Gradual Viper, $\sigma \in \Sigma$ is used to denote a symbolic state, which is a 6-tuple given by ($\isimp$, $h_?$, $h$, $\gamma$, $\pi$, $\rchecks$) where $\isimp$ is a Boolean denoting if the state is imprecise, $h_?$ is the optimistic heap containing the optimistically assumed heap chunks for accessibility predicates, $h$ is the symbolic heap, $\gamma$ is the symbolic store, $\pi$ is the path condition,  and $\rchecks$ is the collection of the runtime checks. Q denotes the remainder continuation which is the remaining symbolic execution that needs to be performed. 

Each pure function in a Gradual Viper program is checked for well-formedness alongside with the methods and predicates. In Figure 3, well-formedness rule for functions in Gradual Viper is given. Well-formedness check for functions in Viper has been modified for Gradual Viper. In Figure 4, well-formed formula function definitions are given. Well-formed formula function definitions have been modified for checking the well-formedness of function postconditions. Since the function postconditions are expressions and not assertions, eval must be used with the function postconditions instead of produce.

\label{sec:figure-draft}

\begin{figure}[htbp]
\footnotesize\ttfamily
\disableTttResize
\hspace{-15.2em} $ \eval{\sigma}{e}{Q} : \Sigma \longrightarrow EXPR \longrightarrow (\Sigma \longrightarrow SVALUE \longrightarrow RESULT) \longrightarrow RESULT$
\begin{alignat*}{2}  
  &\eval{\sigma_1}{func(\overline{e})}{Q} 
    &&= \heval{\sigma_1}{\overline{e}}{(\lambda \sigma_2, \overline{e'} . \\
      &&&\qquad \mlightblue{\rchecks' \assign \sigma_2.\rchecks\{ \origin \assign (\sigma_2,~\sCall{\overline{z}}{func}{\overline{e}},~ \overline{e'}) \}} \\
    &&&\qquad \hconsume{\sigma_2\{\mlightblue{\rchecks \assign \rchecks'}\}}{func_{pre}[\overline{x \mapsto e'}]}{(\lambda \sigma_3, s. \\
    &&&\qquad \mlightyellow{\text{if}\;\; (\text{is-imp}(func)) \;\; \text{then}} \\
    &&&\qquad \quad \mlightyellow{ \sigma_4 \assign \sigma_3\{ h \assign \sigma_2.h, \oh \assign \sigma_2.\oh, \isimp \assign \text{true} \} }\\
    &&&\qquad \mlightyellow{\text{else}} \;\; \sigma_4 \assign \sigma_3\{ h \assign \sigma_2.h, \oh \assign \sigma_2.\oh, \isimp \assign \sigma_2.\isimp \} \\
    &&& \qquad Q(\sigma_4\{\mlightblue{\rchecks \assign \sigma_4.\rchecks\{ \origin \assign \none \}}, func(\overline{e'}, s)})})) \\
    &&& \hspace{-8em} \text{where $func_{pre}$ does not transitively contain another application of $func$} 
\end{alignat*}

\fbox{\begin{tabular}{llll}
\textcolor{light-yellow}{$\blacksquare$} & \small{Handles imprecision} &
\textcolor{light-blue}{$\blacksquare$} & \small{Handles run-time check generation and collection}
\end{tabular}}

\caption{Rule for symbolically executing a pure function in a \gviper program}
\label{fig:figure-draft}
\end{figure}

\newpage

In Figure 5, the eval rule for functions in Gradual Viper is given in continuation-passing style. Yellow and blue colors denote the differences from the eval rule for functions in Viper. Yellow highlighted eval and consume denote the Gradual Viper versions of the CPS rules for eval and consume rather than the Viper rules. Blue highlighted part ensures that where the runtime checks are originated from is tracked.

\end{document}